\documentstyle[preprint,aps]{revtex}
\draft
\title{ac-Field-Controlled Anderson Localization in Disordered
Semiconductor Superlattices}
\author{Martin Holthaus and Gerald H. Ristow}
\address{Fachbereich Physik der Philipps-Universit\"at,
    Renthof 6, D-35032 Marburg, Germany}
\author{Daniel W.~Hone}
\address{Institute for Theoretical Physics,
    University of California at Santa Barbara,
    Santa Barbara, CA 93106}
\date{\today}
\narrowtext
\begin{document}
\maketitle
\begin{abstract}
An ac field, tuned exactly to resonance with the Stark ladder in an ideal
tight binding lattice under strong dc bias, counteracts Wannier-Stark
localization and leads to the emergence of extended Floquet states.
If there is random disorder, these states localize. The localization lengths
depend non-monotonically on the ac field amplitude and become essentially
zero at certain parameters. This effect is of possible relevance for
characterizing the quality of superlattice samples, and for performing
experiments on Anderson localization in systems with well-defined disorder.
\end{abstract}
\pacs{72.15.Rn,73.20.Dx,73.40.Gk}

A major motivation for the introduction of semiconductor superlattices
by Esaki and Tsu~\cite{EsakiTsu70} was the possibility of observing Bloch
oscillations in such effectively one-dimensional periodic structures.
The time of one Bloch oscillation is proportional to the inverse lattice
period, and can therefore become shorter than the typical dephasing times
in superlattices with a spatial period of the order of 100~\AA.

For realizing Bloch oscillations, superlattices of high quality are essential.
However, a certain degree of disorder in these artificial lattices is
inevitable. It is well known that in one spatial dimension even arbitrarily
weak disorder leads to localization of all electronic
eigenstates~\cite{MottTwose61}. In short superlattices of high quality,
consisting of perhaps 100 periods, the localization lengths will exceed
the length of the whole sample, so that localization effects will be negligible
under normal conditions.

In this Letter we demonstrate that the degree of localization in
one-dimensional disordered tight-binding lattices can be controlled by
external homogeneous ac fields, even to such an extent that for certain
field parameters all electronic eigenstates are entirely localized at
individual sites. This effect has at least two practical applications.
First, it can be exploited to characterize the quality of superlattice
samples. Second, and from a more fundamental point of view, it opens up
a new possibility for the experimental investigation of Anderson
localization~\cite{Anderson58,Thouless74,KramerMacKinnon93}: the strength
of an ac field may be used to manipulate the localization lengths
in intentionally disordered superlattices.

We consider a single-band tight binding model in the presence of both
an ac and a dc field:
\begin{equation}
H(t) = H_{0} + H_{int}(t) + H_{random}	\; ,	\label{HAM}
\end{equation}
where the Hamiltonian $H_{0}$ describes an energy band of width $\Delta$
in an ideal, unperturbed lattice with lattice spacing $d$,
\begin{equation}
H_{0} = -\frac{\Delta}{4}\sum_{\ell}\left( |\ell+1 \rangle \langle \ell|
    \, + \, |\ell \rangle \langle \ell+1| \right)       \; ;
\end{equation}
$|\ell\rangle$ denotes a Wannier state at the $\ell$-th site. Next,
\begin{equation}
H_{int}(t) = ed \left[ F_{st} + F_{L}\cos(\omega t) \right]
    \sum_{\ell} |\ell\rangle \, \ell \, \langle \ell |
\end{equation}
describes the interaction with a homogeneous static field of strength $F_{st}$
and an oscillating field of strength $F_{L}$ and frequency $\omega$,
polarized along the lattice direction.
\begin{equation}
H_{random} = \sum_{\ell} \nu_{\ell} \, |\ell\rangle\langle\ell|
\label{RAN}
\end{equation}
introduces site diagonal disorder~\cite{Anderson58}. The random
energies $\nu_{\ell}$ are distributed according to a certain probability
density $\rho(\nu)$; we use a system of units with $\hbar = 1$.

In the absence of disorder, $\nu_{\ell} = 0$ for all sites $\ell$, the
solutions to the time-dependent Schr\"{o}dinger equation
$i\partial_{t}\psi(t) = H(t)\psi(t)$ are given by the so-called
``accelerated Bloch states'', or Houston states~\cite{Houston40}:
\begin{equation}
\psi_{k}(t) \! = \! \frac{1}{\sqrt{N}}\!\sum_{\ell} |\ell\rangle
    \exp\!\left(\!-iq_{k}(t)\ell d
    -i\!\int_{0}^{t}\!\! d\tau \, E(q_{k}(\tau))\right) ,
\label{HOU}
\end{equation}
where $q_{k}(t) = k - eA(t)$;
$A(t) = -F_{st}t - (F_{L}/\omega)\sin(\omega t)$ is the vector potential
of the electric field,
and $E(k) = -(\Delta/2)\cos(kd)$ is the energy dispersion of the
unperturbed band. $N$ denotes the number of lattice sites; finite size
effects are neglected.

Because the Hamiltonian~(\ref{HAM}) is periodic in time with period
$T= 2\pi/\omega$, there should be a complete set of Floquet states,
i.e., of $T$-periodic eigensolutions $u(t)$ to the equation
\begin{equation}
\left(H(t) - i\partial_{t}\right)u(t) = \varepsilon u(t) \; .
\end{equation}
It was realized by Zak~\cite{Zak93} that for vanishing disorder the
construction of these Floquet states, and the calculation of their
quasienergies $\varepsilon$, from the Houston states~(\ref{HOU}) becomes
particularly transparent if $n\omega = eF_{st}d$ with
$n = 0,1,2,\ldots$, i.e., if the energy of $n$ photons precisely matches
the energy difference $eF_{st}d$ induced between adjacent sites by the static
field. For such an ``$n$-photon-resonance'', one finds Floquet states
\begin{equation}
u_{k}(t) = \psi_{k}(t)\exp(+i\varepsilon(k)t)	\label{FLO}
\end{equation}
and quasienergies
\begin{equation}
\varepsilon(k) =
(-1)^{n}J_{n}\!\left(\frac{eF_{L}d}{\omega}\right) E(k) \; ,
\label{QES}
\end{equation}
where $J_{n}$ denotes the Bessel function of order $n$. These
Floquet states are {\em extended} over all the lattice, and are characterized
by the reciprocal lattice vector $k$ as a good quantum number~\cite{Zak93}.
Remarkably, in the absence of disorder, even a weak resonant ac field fully
counteracts Wannier-Stark
localization~\cite{Wannier60,FukuyamaEtAl73,AvronEtAl77,KriegerIafrate88}
that would result from a strong static field alone.

In the absence of ac fields a single defect in an otherwise ideal lattice
supports a localized energy eigenstate. Similarly, a single defect in a
resonantly
driven lattice gives rise to a localized Floquet state: if only $\nu_{0}$
differs from zero in~(\ref{RAN}), the inverse exponential decay length
$L^{-1}$ of the Floquet state supported by the defect, measured in multiples
of the lattice period $d$, is given approxi\-mately~\cite{HolthausEtAl95} by
\begin{equation}
L^{-1} = -2\ln\left(\sqrt{\frac{4\nu_{0}^{2}}{W^{2}} + 1}
- \left| \frac{2\nu_{0}}{W} \right| \right) \; ,
\label{LOC}
\end{equation}
where $W = \Delta |J_{n}(eF_{L}d/\omega)|$ is the width of the
quasienergy band~(\ref{QES}). Thus, $L$ depends non-monotonically on the
ac field strength $F_{L}$, and even becomes zero when $eF_{L}d/\omega$
approaches a zero of the Bessel function $J_{n}$, for arbitrary defect
strength.

Eq.~(\ref{LOC}) looks exactly like the analogous equation for the decay
rate of a time-independent impurity state in an energy band of width $W$.
Thus, this equation indicates that a quasienergy band~(\ref{QES}) behaves,
to some extent, as if it were an ordinary energy band. If this were true
even in the presence of random disorder, a most interesting possibility would
emerge. It has been known since the pioneering work of
Anderson~\cite{Anderson58} that electronic eigenstates in random lattices
are strongly localized
if the typical disorder strength $\bar{\nu}$ becomes comparable to the
energy band width. If the quasienergy band width now takes over the role
of the energy band width in the presence of ac fields, then the degree
of Anderson localization can be controlled by the ac field amplitude.
Since it is possible to fabricate intentionally disordered semiconductor
superlattices, and even to control the amount of lattice disorder
during the growth process, experiments with intentionally disordered
superlattices in far-infrared laser fields~\cite{GuimaraesEtAl93} could open
up an entirely novel access to the physics of localization
phenomena~\cite{MottTwose61,Anderson58,Thouless74,KramerMacKinnon93}.
The confirmation of the hypothesis that Anderson localization in
one-dimensional lattices can be controlled by ac fields is the key result of
the present Letter.

To quantify the degree of localization, we first employ the averaged inverse
participation ratio $P$~\cite{Wegner80}. The Floquet states $u_{m}(t)$
for a disordered, finite lattice of $N$ sites are expanded with respect
to the Wannier states,
\begin{equation}
u_{m}(t) = \sum_{\ell=1}^{N} c_{\ell}^{(m)}(t) \, |\ell\rangle \; .
\end{equation}
Then $P$ is defined as
\begin{equation}
P = \frac{1}{NT} \sum_{\ell,m=1}^{N} \int_{0}^{T} \! dt \,
|c_{\ell}^{(m)}(t)|^{4}	\; .
\label{IPR}
\end{equation}
(Actually, the time-dependence of the localized states turns out to be
very weak, so that averaging over time becomes superfluous.)
If {\em all} states are entirely localized at individual sites,
$P$ approaches unity, wheras it vanishes as $1/N$ if all states are
extended, $|c_{\ell}^{(m)} | \approx 1/\sqrt{N}$ for all $\ell,m$.

For the numerical computations we employ a lattice of 101 sites.
All following results, except Fig.~5, have been obtained for $n=1$,
i.e., $\omega = eF_{st}d$, and $\Delta/\omega = 1.0$.

First, we choose a square disorder distribution,
$\rho(\nu) = 1/(2\nu_{max})$ for $|\nu| \leq \nu_{max}$, and zero otherwise.
Fig.~1 shows the response of the disordered system to the resonant
ac field, for various disorder strengths $\nu_{max}$: when there is disorder,
a certain minimal amplitude is necessary to destroy Wannier-Stark
localization. In agreement with the criterion originally put forward by
Anderson~\cite{Anderson58}, the crossover from strongly to weakly localized
states occurs when the quasienergy band width of the {\em ideal} system has
become comparable to the disorder strength,
$2\bar{\nu}/\Delta \approx |J_{1}(eF_{L}d/\omega)|$,
where $\bar{\nu} = \nu_{max}/\sqrt{3}$ is the variance of $\rho$.

When the field strength is increased further, such that $eF_{L}d/\omega$
approaches the first positive zero $j_{1,1} = 3.83171$ of $J_{1}$, the
quasienergy band width approaches zero again. This leads to the
anticipated effect: Fig.~2 shows the ac-field induced strong Anderson
localization near $j_{1,1}$. Since $P$ almost reaches unity, the Floquet
states become localized essentially at individual sites.

The parameters considered here are not unrealistic for experiments with
far-infrared radiation on semiconductor superlattices~\cite{GuimaraesEtAl93}.
Assuming a scattering time $\tau = 10^{-12}$ seconds,
and $\omega = 2$~meV, one has $\omega\tau \approx 3 > 1$, necessary for
maintaining phase coherence. For this frequency, and a superlattice
period $d = 100$~\AA, an ac field strength $F_{L} = 10\,000$~V/cm already
yields $eF_{L}d/\omega = 5$. The distinct advantage of working with artificial
superlattices is that one may even predetermine the amount of disorder in the
sample. Thus, it is possible to realize somewhat exotic disorder distributions.
As an example, we choose the singular distribution
$\rho(\nu) = 1/(\pi\nu_{max}\sqrt{1 - (\nu/\nu_{max})^{2}})$
for $|\nu| < \nu_{max}$, and zero otherwise. As an alternative measure for
the degree of localization, we compute the variances $\sigma^{(m)}$
of the distributions $p^{(m)}(\ell) = | c_{\ell}^{(m)}(0) |^{2}$
of the Floquet states over the lattice sites, and plot in Fig.~3
the average value $\sigma_{mean}$, again as a function of the scaled ac field
strength $eF_{L}d/\omega$. If all Floquet states were uniformly extended
over all $101$ sites, $\sigma_{mean}$ would be $29.15$.
For weak disorder the numerical data almost reach this value between the
zeros of $J_{1}$. But close to the zeros, there is again practically complete
localization at individual sites. For strong disorder the states do not
recover from their localization, since the quasienergy band does not
become wide enough again.

Fig.~4 shows the corresponding quasienergies for $\nu_{max}/\Delta = 0.2$.
The peculiar clustering of eigenvalues into three groups is caused by the
singular density $\rho(\nu)$; it is not found for the square
distribution~\cite{HolthausEtAl95}. At the zeros of $J_{1}$ the quasienergy
band of the disordered lattice has a width of the order of $2\nu_{max}$.
This shows that Anderson localization in disordered lattices is entirely
different from the ``dynamic localization'' discovered by
Dunlap and Kenkre~\cite{DunlapKenkre86} in ideal, ac-driven
lattices. If one forms a wave packet that is initially localized at an
arbitrary site of an ideal lattice, $H_{random} \equiv 0$, then the wave
packet will remain localized at that site if $eF_{st}d = n\omega$ and,
simultaneously, $eF_{L}d/\omega$ coincides with a zero of $J_{n}$: since the
quasienergies of all the wave packet's components are then equal, all
components acquire precisely the same phase factor during one cycle of the ac
field, so that the wave function simply reproduces itself, apart from an
overall phase factor~\cite{HolthausHone93}. However, the Floquet
states~(\ref{FLO}) remain extended over all the lattice. Dynamic localization
of wave packets in ideal tight binding lattices merely reflects
the degeneracy of all quasienergies at the zeros of $J_{n}$. It has been shown
recently that this effect persists even in the presence of Coulomb
interactions~\cite{MeierEtAl94}.

In contrast, ac-field induced strong Anderson localization in disordered
lattices implies the localization {\em of the Floquet states themselves},
cf. Figs.~2,3, and there is {\em no} total degeneracy of the eigenvalues,
cf. Fig.~4. Experimentally, there is a clear-cut signature for ac-field
induced Anderson localization: if all the eigenstates are localized, the only
mechanism enabling conductance will be variable range hopping. At the zeros
of $J_{n}$, the conductance of a disordered superlattice should therefore
{\em decrease} with decreasing temperature, whereas it should {\em increase}
in between, where phonons impede transport via the (effectively) extended
states.

We reemphasize the special role of {\em resonant} ac fields,
$eF_{st}d = n\omega$~\cite{Zak93}. If this condition is not satisfied, the
states remain Wannier-Stark localized; there are no ``extended'' states at
all. Fig.~5 shows the quasienergy spectrum for precisely the same
realization of disorder as employed in Fig.~4, but for
$eF_{st}d = 1.11 \, \omega$. (In the ideal, non-resonant system, the
quasienergies are simply $\varepsilon_{\ell} = eF_{st}\ell d \bmod \omega$.)
The corresponding values of $\sigma_{mean}$ stay below $2.0$ in the entire
range of $eF_{L}d/\omega$.

Up to now, interest in the manipulation of Anderson localization
by external fields remained restricted to magnetic fields~\cite{Bergmann84}.
The present results indicate that combined dc- and ac electric fields also
have pronounced and systematic effects on localization phenomena, and that
these effects can be studied with the help of existing
facilities~\cite{GuimaraesEtAl93} in disordered semiconductor superlattices
-- with the distinct advantage that even the sample-specific realizations of
disorder are under experimental control.

We should like to acknowledge most valuable discussions with
S.J. Allen, F. Gebhard, S. Grossmann, and P. Thomas. One of us (D.H.)
was supported by NSF Grant No.\ PHY 94-07194.

\begin{figure}
\caption[FIG.~1] {Destruction of Wannier-Stark localization by a resonant
    ac field, $eF_{st}d = \omega$: inverse participation ratios~(\ref{IPR})
    as functions of the scaled ac field strength. The lattice has 101 sites;
    $\Delta/\omega = 1.0$. The disorder distribution is
    $\rho(\nu) = 1/(2\nu_{max})$ for $|\nu| \leq \nu_{max}$,
    with $\nu_{max}/\Delta = 0.10$, $0.05$, $0.02$, and $0.01$
    (top to bottom). Indicated are the field strengths
    where $2\bar{\nu}/\Delta = J_{1}(eF_{L}d/\omega)$,
    with $\bar{\nu} = \nu_{max}/\sqrt{3}$.}
\end{figure}

\begin{figure}
\caption[FIG.~2] {Strong Anderson localization occurs when $eF_{L}d/\omega$
    approaches 3.83171, the first positive zero of $J_{1}$. Other parameters
    are as in Fig.~1.}
\end{figure}

\begin{figure}
\caption[FIG.~3] {Average variance $\sigma_{mean}$ of the spatial distribution
    of the Floquet states in units of the lattice spacing. The disorder
    distribution is
    $\rho(\nu) = 1/(\pi\nu_{max}\sqrt{1 - (\nu/\nu_{max})^{2}})$
    for $|\nu| < \nu_{max}$, with
    $\nu_{max}/\Delta = 0.05$, $0.10$, $0.15$, and $0.20$ (top to bottom).
    Note $\sigma_{mean} = 29.15$ for uniformly extended states.}
\end{figure}

\begin{figure}
\caption[FIG.~4] {Quasienergy spectrum for the same disorder distribution
    as used in Fig.~3, and $\nu_{max}/\Delta = 0.2$.  The arrows indicate the
    first two positive zeros of $J_{1}$.}
\end{figure}

\begin{figure}
\caption[FIG.~5] {Quasienergy spectrum for a non-resonant case,
    $eF_{st}d = 1.11\,\omega$. $\rho(\nu)$ is the same as in Fig.~4.}
\end{figure}

\end{document}